\documentclass[pra,aps,amsmath,amssymb,showkeys,showpacs]{revtex4}
\usepackage{graphicx}

\newcommand{\cla}{\mathcal{A}}
\newcommand{\clb}{\mathcal{B}}
\newcommand{\cle}{\mathcal{E}}
\newcommand{\pen}{\openone}
\newcommand{\bnil}{\mathbf{0}}
\newcommand{\iu}{\mathtt{i}}
\newcommand{\xdif}{\mathrm{d}}
\newcommand{\bro}{\boldsymbol{\rho}}
\newcommand{\bsg}{\boldsymbol{\sigma}}

\newcommand{\blm}{\boldsymbol{\lambda}}
\newcommand{\hm}{\mathsf{H}}
\newcommand{\am}{\mathsf{A}}
\newcommand{\ems}{\mathsf{E}}

\newcommand{\um}{\mathsf{U}}

\newcommand{\xm}{\mathsf{X}}
\newcommand{\sm}{\mathsf{S}}
\newcommand{\nam}{\mathsf{N}}
\newcommand{\bsl}{\boldsymbol{L}}
\newcommand{\pisf}{\mathsf{\Lambda}}
\newcommand{\mpi}{\mathsf{\Pi}}

\newcommand{\Tr}{\mathrm{Tr}}

\newcommand{\hw}{\widetilde{H}}
\newcommand{\dw}{\widetilde{\Delta}}

\begin{document}
\clearpage
\preprint{}

\title{On generalized entropies and information-theoretic Bell inequalities under decoherence}

\author{Alexey E. Rastegin}
\email{rast@api.isu.ru; alexrastegin@mail.ru}
\affiliation{Department of Theoretical Physics, Irkutsk State University,
Gagarin Bv. 20, Irkutsk 664003, Russia}

\begin{abstract}
We consider information-theoretic inequalities of the Bell type in
the presence of decoherence. It is natural that too strong
coupling with the environment can prevent an observation of quantum
correlations. In this regard, the use of various entropic
functions may give additional capabilities to reveal desired
correlations. It was already shown that the Bell and Leggett--Garg
inequalities in terms conditional Tsallis entropies are more
sensitive in the cases of detection inefficiencies. In this paper,
we study capabilities of generalized conditional entropies of the
Tsallis type in analyzing the Bell theorem in
decoherence scenarios. Two forms of the conditional Tsallis $q$-entropy are
known in the literature. We show that each of them can be used for
defining a metric in the probability space of interest. Such
metrics can be used in realizing the so-called triangle principle.
The triangle principle has recently been proposed as a unifying
approach to questions of local realism and non-contextuality.
Applying the triangle principle leads to the two families of
$q$-metric inequalities of the Bell type.
Information-theoretic formulations in terms of the $q$-entropic
metrics are first discussed for the CHSH scenario in dephasing
environment. Then we also revisit $q$-entropic inequalities of the
Leggett--Garg type. An environmental influence is modeled by  the
phase damping channel and by the depolarizing channel.
\end{abstract}


\maketitle

\section{Introduction}\label{sec1}

Non-classical nature of quantum correlations was independently
emphasized in the Schr\"{o}dinger ``cat paradox'' paper
\cite{cat35} and in the Einstein--Podolsky--Rosen paper
\cite{epr35}. This character is clearly manifested in some
experiments such as Bohm's version of the EPR argument
\cite{bohm51}. Correlations observed experimentally are related to
statistical predictions and probability distributions
\cite{maleeh08}. As was shown in the seminal papers by Bell
\cite{bell64,bell66} and by Kochen and Specker \cite{ks67},
quantum mechanics is not consistent with some assumptions based on
a classical experience. Bell's ideas have allowed to recast the
problem of hidden variables as an experimentally tested statement
\cite{az99}. Leggett--Garg inequalities \cite{lg85} form a
direction inspired by the Bell theorem. Such relations are based
on the two assumptions known as the macroscopic realism and the
noninvasive measurability at the macroscopic level \cite{eln13}.
As was shown by Bell, predictions of quantum theory is not
compatible with the assumption of local realism. Similarly, the
Kochen--Specker theorem and the Leggett--Garg inequalities stated
that quantum mechanics is incompatible with the assumptions of
non-contextuality and macrorealism, respectively. Since
Leggett--Garg inequalities probe correlations of a single system
measured at different times, decoherence is one of crucial
problems in practice. Violations of the Leggett--Garg
inequalities under decoherence were experimentally studied in
Refs. \cite{pal10,xu11}.

Original Bell inequalities were written as a restriction on mean
values \cite{bell64}. The Greenberger--Horne--Zeilinger approach
has given a statement without inequalities \cite{ghsz90}.
Formulations of Bell inequalities in terms of mean values
typically assumes a fixed number of observable outcomes. Entropic
treatment allows a unified expression irrespectively to a number
of outcomes \cite{rchtf12}. To test the local realism
experimentally, several scenarios are known. The
Clauser--Horne--Shimony--Holt (CHSH) scenario \cite{chsh69} is
probably the most known setup. Entropic versions of Bell's theorem
were considered in Refs. \cite{BC88,cerf97}. These papers were
mainly focused on the CHSH scenario. The
Klyachko--Can--Binicio\v{g}lu--Shumovsky (KCBS) scenario
\cite{kly08} is currently the subject of active research. The CHSH
and KCBS scenarios are respectively the $n=4$ and $n=5$ cases of
the $n$-cycle scenario \cite{lsw11,aqbcc12}. For the $n$-cycle,
the quantum violations occur for all $n$, though technical
questions make their observation harder for $n\gg1$
\cite{bcdfa13}. Information-theoretic Bell inequalities for the
KCBS scenario were examined in Refs. \cite{rchtf12,krk12}.  For
both the CHSH and KCBS scenarios, inequalities in terms of Tsallis
$q$-entropies were studied in Ref. \cite{rastqic14}. The
Leggett--Garg case deals with a cycle of observables taken at different
times. In Ref. \cite{pkdk13}, the triangle principle has been
proposed as a new approach to the non-locality and contextuality.
Similar ideas were considered in Ref. \cite{santos14}.
Applications of this principle to qutrits with use of the
Tsallis-type metrics were recently examined in Ref. \cite{mkk14}.

In generalized Bell scenarios, we deal with the problem of
deciding, whether observed data is compatible with a presumed
causal relation between the variables \cite{clg2014}. It
traditionally focuses on settings, when the region of compatible
observations corresponds to some convex polytope. In principle,
such polytopes can be represented by finitely many Bell
inequalities. However, the size of characterization grows very
fast as number and/or dimensionality of involved observables
increases. For instance, a complete description  of the $n$-cycle
is given by an exponential number of tight inequalities
\cite{aqbcc12}. Entropic inequalities are able to describe
extended Bell scenarios that define complicated non-convex sets in
the probabilistic space of interest
\cite{clg2014,hlp2014,cmg2014}. Entropic inequalities of the Bell
type are attractive due to their capabilities in studies of
setting with arbitrary number of outcomes and inefficiencies of
measurement devices. At the same time, entropic inequalities give
only sufficient conditions of the non-locality or contextuality
\cite{rchtf12}. There are probability distributions that do
violate Bell's inequality and do not its entropic counterpart. As
was shown in Ref. \cite{rastqic14}, use of generalized entropies
allows to extend a class of probability distributions, whose
non-locality or contextuality can be expressed by an entropic
approach. It is an alternative to the following approach. Adding a
shared randomness \cite{rch13}, inequalities with the standard
Shannon entropies can sometimes be turned into a necessary and
sufficient condition. This has been shown for the $n$-cycle with
dichotomic outcomes \cite{rch13} in noise-free and error-free
settings. For more outcomes or decoherence scenarios, inequalities
with $q$-entropies are appropriate.

The aim of the present work is to study information-theoretic Bell
inequalities based on generalized conditional entropies. Some
advantages of this approach were already examined 
\cite{rastqic14,mkk14,rastq14}. We will mainly focus on entropic
inequalities of the Bell type under decoherence. This question
seems to be not addressed in the literature. Indeed, variations of
the parameter in $q$-entropic inequalities are useful in analyzing
cases with detection inefficiencies \cite{rastqic14,rastq14}.
Formulation of restrictions of the Leggett--Garg type in terms of
the Shannon entropies was examined in Ref. \cite{uksr12}. A
$q$-entropic extension of this question has been discussed in Ref.
\cite{rastq14}. We also aim to study information-theoretic Bell
inequalities in the presence of decoherence. The contribution of
the present paper is two-fold. First, we show that each of the two
known forms of conditional $q$-entropy leads to the corresponding
metric between random variables. One of the conditional
$q$-entropies obeys the chain rule \cite{sf06}, whence the
triangle inequality for a metric follows for $q\geq1$. However,
other conditional $q$-entropy does not share the chain rule. It is
not obvious that a legitimate metric could be obtained in this
way. Second, we consider violation of $q$-metric inequalities of
the Bell type under decoherence. Decohering processes are one of
crucial problems for an observation of quantum correlations in 
practice. In particular, dephasing processes can prefer a
detection of such correlations.

In this paper, we will show that $q$-entropic inequalities could
be useful in analysis of data of experiments in decohering
environment. The paper is organized as follows. In Section
\ref{sec2}, we consider those metrics that can be based on the
conditional $q$-entropies. It is shown that the known conditional
forms of the Tsallis entropy both lead to a legitimate metric for
$q\geq1$. Here, the triangle inequality is most important from the
viewpoint of applications of the triangle principle. For one of
the cases considered, the triangle inequality directly follows
from the chain rule. In the second case, the desired result is
obtained due to independent reasons. In Section \ref{sec3},
inequalities of the Bell type are written as $q$-metric
inequalities for the CHSH scenario with noise.  Section \ref{sec4}
is devoted to $q$-metric Leggett--Garg inequalities under
decoherence. We demonstrate advantages of metric inequalities with
some parameter that can be varied for maximizing a desired
violation. Varying the parameter in $q$-entropic inequalities, a
violation of the restrictions considered may become much more
robust to decoherence. As models of quantum noise, the phase
damping and depolarizing channels are utilized. In Section
\ref{sec5}, we conclude the paper with a summary of results.

\section{Conditional Tsallis entropies and related metrics}\label{sec2}

In this section, we discuss required properties of the
$q$-entropies and their conditional forms. Two kinds of the
$q$-entropic metric will be examined. Let discrete random variable
$X$ take values on a finite set $\Omega_{X}$ of cardinality
$\#\Omega_{X}$. The non-extensive entropy of degree $q>0\neq1$ is
defined by \cite{tsallis}
\begin{equation}
H_{q}(X):=\frac{1}{1-q}{\,}{\left(\sum_{{\,}x\in\Omega_{X}} p(x)^{q}
- 1\right)}
{\,}. \label{tsaent}
\end{equation}
With the factor $\left(2^{1-q}-1\right)^{-1}$ instead of
$(1-q)^{-1}$, this function was examined by Havrda and Charv\'{a}t
\cite{havrda} and later by Dar\'{o}czy \cite{ZD70}. In statistical
physics, the entropy (\ref{tsaent}) is extensively used due to
Tsallis \cite{tsallis}.

Obviously, the entropy (\ref{tsaent}) is concave for all $q>0$. It
is convenient to rewrite (\ref{tsaent}) as
\begin{equation}
H_{q}(X)=-\sum_{x\in\Omega_{X}}p(x)^{q}{\,}\ln_{q}{p}(x)
=\sum_{x\in\Omega_{X}}p(x){\>\,}{\ln_{q}}{\left(\frac{1}{p(x)}\right)}
\ . \label{tsaln}
\end{equation}
Here, we used the $q$-logarithm defined for $q>0\not=1$ and
$\xi>0$ as
\begin{equation}
\ln_{q}(\xi)=\frac{\xi^{1-q}-1}{1-q}
\ . \label{lnadf}
\end{equation}
In the limit $q\to1$, we obtain $\ln_{q}(\xi)\to\ln{\xi}$ and the
standard Shannon entropy
\begin{equation}
H_{1}(X)=-\sum_{x\in\Omega_{X}}p(x){\,}\ln{p}(x)
\ . \label{shaln}
\end{equation}
For the uniform distribution, the maximal value
${\ln_{q}}{(\#\Omega_{X})}$ of (\ref{tsaent}) is reached. The
R\'{e}nyi entropies \cite{renyi61} form another especially
important family of one-parametric extensions of the Shannon
entropy. The R\'{e}nyi entropies are beyond the scope of the present
work. Some properties and applications of such entropies are
discussed in the book \cite{bengtsson}.

To define a metric, we will use conditional entropies. For
brevity, we will omit symbols $\Omega_{X}$ and $\Omega_{Y}$ in
entropic sums. The standard conditional entropy is defined by
\cite{CT91}
\begin{equation}
H_{1}(X|Y):=\sum\nolimits_{y} p(y){\,}H_{1}(X|y)
=-\sum\nolimits_{x}\sum\nolimits_{y} p(x,y){\,}\ln{p}(x|y)
\ . \label{cshen}
\end{equation}
Here, we use Bayes' rule $p(x|y)=p(x,y)/p(y)$ and the particular
function
\begin{equation}
H_{1}(X|y)=-\sum\nolimits_{x}p(x|y){\,}\ln{p}(x|y)
\ . \label{csheny}
\end{equation}
In the literature, two kinds of the conditional $q$-entropy were
discussed \cite{sf06}. These forms are respectively inspired by
the two expressions, which are shown in (\ref{tsaln}). The first
form is defined as \cite{sf06}
\begin{equation}
H_{q}(X|Y):=\sum\nolimits_{y} p(y)^{q}{\,}H_{q}(X|y)
\ , \label{hct1}
\end{equation}
where
\begin{equation}
H_{q}(X|y):=\frac{1}{1-q}{\,}\left(\sum\nolimits_{x} p(x|y)^{q}-1\right)
{\,}. \label{hctm0}
\end{equation}
Similarly to (\ref{tsaln}), the equivalent expressions are written
as
\begin{align}
H_{q}(X|y)&=-\sum\nolimits_{x} p(x|y)^{q}{\,}\ln_{q}{p(x|y)}
\label{hct0}\\
&=\sum\nolimits_{x} p(x|y){\>}{\ln_{q}}{\left(\frac{1}{p(x|y)}\right)}
{\,}. \label{hct00}
\end{align}
The conditional entropy (\ref{hct1}) is, up to a factor, the
quantity introduced by Dar\'{o}czy \cite{ZD70}. For all $q>0$, we
have the chain rule \cite{ZD70,sf06}
\begin{equation}
H_{q}(X,Y)=H_{q}(Y|X)+H_{q}(X)
=H_{q}(X|Y)+H_{q}(Y)
\ . \label{chrl}
\end{equation}
With $q=1$, we have the chain rule with the standard
conditional entropy (\ref{cshen}). An immediate extension of
(\ref{chrl}) for more than two random variables was given in Ref.
\cite{sf06}. Relations of such kind play an important role in
many information-theoretic derivations. For instance, the
Braunstein--Caves derivation \cite{BC88} of entropic Bell
inequalities is based on the chain rule for the Shannon entropy.

As was noted in Ref. \cite{rchtf12}, information-theoretic Bell
inequalities for the $n$-cycle scenario can be represented in
terms of the mutual information. Similarly to the standard case,
the mutual $q$-information can be defined as \cite{sf06}
\begin{equation}
I_{q}(X;Y):=H_{q}(X)-H_{q}(X|Y)
=H_{q}(Y)-H_{q}(Y|X)
\ . \label{minq}
\end{equation}
The equivalence of the two last expressions is provided by the
chain rule (\ref{chrl}). Formulation of the Bell theorem in terms
of the mutual $q$-information has been addressed in Ref.
\cite{rastqic14}.

Using the particular functional (\ref{hctm0}), the second form of
conditional $q$-entropy is introduced as \cite{sf06}
\begin{equation}
\hw_{q}(X|Y):=\sum\nolimits_{y}p(y){\,}H_{q}(X|y)
\ . \label{hct2}
\end{equation}
Note that this form of conditional entropy does not share the
chain rule of usual kind \cite{sf06}. Hence, it is not directly
related to the mutual $q$-information. Nevertheless, the entropy
(\ref{hct2}) has found to be useful at least as an auxiliary
quantity \cite{sf06,rastkyb}. The conditional entropy (\ref{hct2})
can also be used for measuring a distance between random
variables.

The standard conditional entropy leads to the following metric
\cite{hrb73}:
\begin{equation}
\Delta_{1}(X,Y):=H_{1}(X|Y)+H_{1}(Y|X)
\ . \label{d1xy}
\end{equation}
General properties of information distances are considered in Ref.
\cite{benn98}. The author of Ref. \cite{sf06} discussed three
forms of an entropic distance based on the Tsallis entropies.
First of these distances is defined similarly to (\ref{d1xy}):
\begin{equation}
\Delta_{q}(X,Y):=H_{q}(X|Y)+H_{q}(Y|X)
\ . \label{dqxy}
\end{equation}
Due to (\ref{minq}), we can recast (\ref{dqxy}) as
$\Delta_{q}(X,Y)=H_{q}(X,Y)-I_{q}(X;Y)$. As was shown in Ref.
\cite{sf06}, the quantity (\ref{dqxy}) is a metric for $q\geq1$.
It satisfies the following properties.
\begin{itemize}
  \item[(i)]{$\Delta_{q}(X,Y)\geq0$ (non-negativity);}
  \item[(ii)]{$\Delta_{q}(X,Y)=0$ if and only if $X=Y$ (identity axiom);}
  \item[(iii)]{$\Delta_{q}(X,Y)=\Delta_{q}(Y,X)$ (symmetry);}
  \item[(iv)]{$\Delta_{q}(X,Z)\leq\Delta_{q}(X,Y)+\Delta_{q}(Y,Z)$ (triangle inequality).}
\end{itemize}
The last property is easily derived from the inequality
\cite{sf06}
\begin{equation}
H_{q}(X|Z)\leq{H}_{q}(X|Y)+H_{q}(Y|Z)
\ , \label{trhq}
\end{equation}
which holds for $q\geq1$. Other $q$-entropic metrics are defined
in terms of correlation coefficients \cite{sf06}. One form of
correlation coefficients is introduced as the ratio of the mutual
$q$-information to the joint $q$-entropy. Then difference between
$1$ and this correlation coefficient leads to a metric for $q\geq1$
\cite{sf06}. It can also be interpreted as the result of division
of (\ref{dqxy}) by the joint entropy $H_{q}(X,Y)$. Another
correlation coefficient is defined as the ratio of the mutual
$q$-information to the maximum of the entropies $H_{q}(X)$ and
$H_{q}(Y)$. Hence, one leads to the third distance considered in
Ref. \cite{sf06}. It should be emphasized that the mentioned
quantities are metrics only for $q\geq1$. Further, validity of the
triangle inequality for these distances is closely related to the
chain rule. In Ref. \cite{mkk14}, the mentioned $q$-entropic
metrics were used to study Bell inequalities for a pair of
entangled qutrits. These metrics can all be represented in terms
of the mutual $q$-information together with either $H_{q}(X,Y)$ or
$\max\bigl\{H_{q}(X),H_{q}(Y)\bigr\}$.

We shall now examine a $q$-entropic distance which cannot be
expressed in terms of the mutual $q$-information. The conditional
$q$-entropy (\ref{hct2}) does not share the chain rule.
Nevertheless, this conditional form leads to a legitimate metric
as well. We shall analyze the question in more detail, since it
seems to be not addressed in the literature. Similarly to
(\ref{dqxy}), we can introduce another quantity
\begin{equation}
\dw_{q}(X,Y):=\hw_{q}(X|Y)+\hw_{q}(Y|X)
\ . \label{dwqxy}
\end{equation}
It is easy to check that the properties (i)--(iii) remain valid
for (\ref{dwqxy}). The only question concerns the triangle
inequality. To resolve the question, we will examine some
essential properties of the entropy (\ref{hct2}).

\newtheorem{t21}{Proposition}
\begin{t21}\label{tem1}
For $q>0$, the conditional entropy (\ref{hct2}) satisfies
\begin{equation}
\hw_{q}(X,Y|Z)\geq\hw_{q}(X|Z)
\ . \label{rtem1}
\end{equation}
\end{t21}

{\bf Proof.} Since the standard case $q=1$ is well known, we
further assume $q\neq1$. Let positive numbers $a(y)$ satisfy
$\sum_{y}a(y)=1$. We then have
\begin{align}
\sum\nolimits_{y}a(y)^{q}&\geq1
&(0<q<1)
\ , \label{cs190}\\
\sum\nolimits_{y}a(y)^{q}&\leq1
&(1<q<\infty)
\ . \label{cs191}
\end{align}
Combining these relations with $\sum_{y}p(x,y|z)=p(x|z)$, we
obtain
\begin{align}
\sum_{y}{\Bigl(p(x,y|z)^{q}-p(x,y|z)\Bigr)}
&\geq{p}(x|z)^{q}-p(x|z)
&(0<q<1)
\ , \label{pqpy0}\\
\sum_{y}{\Bigl(p(x,y|z)^{q}-p(x,y|z)\Bigr)}
&\leq{p}(x|z)^{q}-p(x|z)
&(1<q<\infty)
\ . \label{pqpy1}
\end{align}
Summarizing with respect to $x$ and taking the sign of the factor
$(1-q)^{-1}$, we have arrived at a conclusion. For all $q>0\neq1$,
one gives
\begin{equation}
H_{q}(X,Y|z)\geq{H}_{q}(X|z)
\ . \label{pqpy11}
\end{equation}
Multiplying (\ref{pqpy11}) by $p(z)$ and summing with respect to
$z$, we finally obtain (\ref{rtem1}). $\blacksquare$

It is clear that the result (\ref{rtem1}) can be generalized as
follows. For real $q>0$ and integer $n\geq1$, we have
\begin{equation}
\hw_{q}(X_{1},\ldots,X_{n},X_{n+1}|Z)
\geq\hw_{q}(X_{1},\ldots,X_{n}|Z)
\ . \label{rtem1n}
\end{equation}
We refrain from presenting details of the argumentation. The
relations (\ref{rtem1}) and (\ref{rtem1n}) will be used below in
deriving the triangle inequality. As was already mentioned, the
conditional entropy (\ref{hct2}) does not share the chain rule
\cite{sf06}. Instead, we will use another statement.

\newtheorem{t22}[t21]{Proposition}
\begin{t22}\label{tem2}
The conditional entropy (\ref{hct2}) satisfies the following
inequalities:
\begin{align}
\hw_{q}(X,Y|Z)-\hw_{q}(Y|Z)&\geq\hw_{q}(X|Y,Z)
&(0<q<1)
\ , \label{rtem20}\\
\hw_{q}(X,Y|Z)-\hw_{q}(Y|Z)&\leq\hw_{q}(X|Y,Z)
&(1<q<\infty)
\ . \label{rtem21}
\end{align}
\end{t22}

{\bf Proof.} Using $p(x,y|z)/p(y|z)=p(x|y,z)$ and the definition
(\ref{hctm0}), we merely write
\begin{align}
&H_{q}(X,Y|z)-H_{q}(Y|z)=\frac{1}{1-q}
\left(\sum\nolimits_{xy}p(x,y|z)^{q}-\sum\nolimits_{y}p(y|z)^{q}\right)
\nonumber\\
&=\sum\nolimits_{y}p(y|z)^{q}{\>}
\frac{1}{1-q}{\,}\left(\sum\nolimits_{x} p(x|y,z)^{q}-1\right)
=\sum\nolimits_{y}p(y|z)^{q}{\,}H_{q}(X|y,z)
\ . \label{hqhqz}
\end{align}
As $p(y|z)\leq1$, replacing $p(y|z)^{q}$ with $p(y|z)$ leads to
\begin{align}
H_{q}(X,Y|z)-H_{q}(Y|z)&\geq\sum\nolimits_{y} p(y|z){\,}H_{q}(X|y,z)
&(0<q<1)
\ , \label{hqhqz0}\\
H_{q}(X,Y|z)-H_{q}(Y|z)&\leq\sum\nolimits_{y} p(y|z){\,}H_{q}(X|y,z)
&(1<q<\infty)
\ . \label{hqhqz1}
\end{align}
Further, we note $p(z){\,}p(y|z)=p(y,z)$. Multiplying
(\ref{hqhqz0}) and (\ref{hqhqz1}) by $p(z)$ and summing with
respect to $z$, we complete the proof. $\blacksquare$

In principle, the formula (\ref{rtem21}) can be regarded as a weak
version of the chain rule. Note that the standard conditional
entropy (\ref{cshen}) obeys the equality
\begin{equation}
H_{1}(X,Y|Z)-H_{1}(Y|Z)=H_{1}(X|Y,Z)
\ . \label{rtm200}
\end{equation}
We can obtain (\ref{rtm200}) by taking the limit $q\to1$ in both
the relations (\ref{rtem20}) and (\ref{rtem21}). We are now ready
to prove that the conditional entropy (\ref{hct2}) of degree
$q\geq1$ obeys the triangle inequality. This result is formulated
as follows.

\newtheorem{t23}[t21]{Proposition}
\begin{t23}\label{tem3}
For $q\geq1$, the conditional entropy (\ref{hct2}) satisfies the
triangle inequality
\begin{equation}
\hw_{q}(X|Z)\leq\hw_{q}(X|Y)+\hw_{q}(Y|Z)
\ . \label{trhwq}
\end{equation}
\end{t23}

{\bf Proof.} Using the properties (\ref{rtem1}) and
(\ref{rtem21}), we obtain
\begin{equation}
\hw_{q}(X|Z)\leq\hw_{q}(X,Y|Z)\leq\hw_{q}(X|Y,Z)+\hw_{q}(Y|Z)
\ . \label{phwq1}
\end{equation}
The first inequality holds for all $q>0$, whereas the second one
generally holds for $q\geq1$. We further recall the fact that
conditioning on more can only reduce the conditional entropy
(\ref{hct2}). Namely, for all $q>0$ we have \cite{rastit}
\begin{equation}
\hw_{q}(X|Y,Z)\leq\hw_{q}(X|Y)
\ . \label{cnom0}
\end{equation}
Combining (\ref{phwq1}) with (\ref{cnom0}) completes the proof.
$\blacksquare$

By permutations, we also write
$\hw_{q}(Z|X)\leq\hw_{q}(Z|Y)+\hw_{q}(Y|X)$. Adding the latter to
(\ref{trhwq}), for $q\geq1$ we obtain
\begin{equation}
\dw_{q}(X,Z)\leq\dw_{q}(X,Y)+\dw_{q}(Y,Z)
\ . \label{trdwq}
\end{equation}
In other words, for $q\geq1$ the entropic quantity (\ref{dwqxy})
is a legitimate metric. Thus, both the quantities (\ref{dqxy}) and
(\ref{dwqxy}) can be adopted as information distances in realizing
the triangle principle. In the next sections, we will consider
this question with respect to the Bell theorem and the
Leggett--Garg inequalities.

\section{Metric inequalities for the CHSH scenario in dephasing environment}\label{sec3}

In this section, we will study $q$-metric inequalities for the
CHSH scenario with focusing on the role of decoherence. The CHSH
scenario is a primary example of the so-called $n$-cycle scenarios
\cite{lsw11,aqbcc12}. It is typically used in studies of
conceptual questions of quantum theory \cite{unr06,dch09}. The
notion of marginal scenarios provides a general way to treat
related properties of probability distributions
\cite{rchtf12,chfr13}. The triangle principle provides another
general approach to the problem \cite{pkdk13}.

We will formulate quantitative relations in terms of the
$q$-metrics (\ref{dqxy}) and (\ref{dwqxy}) for $q\geq1$. Let us
recall briefly details of the CHSH scenario. In this scenario, we
deal with an entanglement of two spacelike separated subsystems
$\cla$ and $\clb$. Let observables $A$ and $A^{\prime}$ be used
for one subsystem, and let observables $B$ and $B^{\prime}$ be
used for other. No one of the pairs $\{A,A^{\prime}\}$ and
$\{B,B^{\prime}\}$ is jointly measurable. Each element of
$\{A,A^{\prime}\}$ is compatible with each element of
$\{B,B^{\prime}\}$, since these sets are related to different
subsystems. Applying the triangle inequality, we simply obtain two
relations
\begin{align}
\Delta_{q}(A,B)&\leq\Delta_{q}(A,B^{\prime})+\Delta_{q}(B^{\prime},A^{\prime})+\Delta_{q}(A^{\prime},B)
\ , \label{chshdq}\\
\dw_{q}(A,B)&\leq\dw_{q}(A,B^{\prime})+\dw_{q}(B^{\prime},A^{\prime})+\dw_{q}(A^{\prime},B)
\ . \label{chshwq}
\end{align}
In the usual CHSH scenario, each of the observables has two
possible outcomes rescaled as $\pm1$. This assumption leads to
concrete form of the bound on mean values. However, entropic
formulations of Bell's theorem have the same form irrespectively
to the number of outcomes or the chosen scale for observables
\cite{rchtf12}. The authors of Ref. \cite{BC88} derived Bell's
inequality for the CHSH scenario in terms of the Shannon
entropies. This inequality is often referred to as the
Braunstein--Caves inequality.

Following Ref. \cite{BC88}, we consider a quantum spin-$s$ system.
To exemplify violations of the relations (\ref{chshdq}) and
(\ref{chshwq}), one uses the following setup. Two
counter-propagating spin-$s$ particles are emitted by the decay of
a system with zero angular momentum. In the simplest case $s=1/2$,
we use the operator $\sm_{z}=(\hbar/2)\bsg_{z}$ with the
eigenstates
\begin{equation}
|0\rangle=
\begin{pmatrix}
1 \\
0
\end{pmatrix}
\, , \qquad
|1\rangle=
\begin{pmatrix}
0 \\
1
\end{pmatrix}
\, . \label{01bs}
\end{equation}
The state of two particles with zero total momentum is written as
\begin{equation}
|\Phi\rangle=\frac{1}{\sqrt{2}}{\,}
\Bigl(
|0\rangle\otimes|1\rangle-|1\rangle\otimes|0\rangle
\Bigr)
\> . \label{ztsp}
\end{equation}
We now take the four unit vectors $\vec{a}$, $\vec{a}^{\prime}$,
$\vec{b}$, and $\vec{b}^{\prime}$. In the quantum-mechanical
description, the quantities $A$ and $A^{\prime}$ are represented
as the operators $\vec{a}\cdot\vec{\sm}$ and
$\vec{a}^{\prime}\cdot\vec{\sm}$. The quantities $B$ and
$B^{\prime}$ are given in the same way. Following Ref.
\cite{BC88}, we consider four coplanar vectors such that
\begin{equation}
\measuredangle(\vec{a},\vec{b}^{\prime})=\measuredangle(\vec{b}^{\prime},\vec{a}^{\prime})
=\measuredangle(\vec{a}^{\prime},\vec{b})=\theta/3,
\qquad
\measuredangle(\vec{a},\vec{b})=\theta
\ . \label{angls}
\end{equation}
Using of coplanar vectors is easy to realize and widely used. We
will see that such a choice is illustrative for comparing
different entropies in the noisy case. Due to the properties of
the standard conditional entropy, Braunstein and Caves formulated
an information-theoretic inequality of the Bell type \cite{BC88}.
In the considered situation, their result is equivalent to the
case $q=1$ of the formulas (\ref{chshdq}) and (\ref{chshwq}). The
Braunstein--Caves inequality expresses the fact that there exists
some joint probability distribution for the four random variables.
In principle, the relation (\ref{chshdq}) can also be reached on
this ground \cite{rastqic14}. Here, the chain rule (\ref{chrl}) is
very important. Since the conditional entropy (\ref{hct2}) does
not share the chain rule, the relation (\ref{chshwq}) cannot be
obtained in such a way.

It was found that the strength of violation of the
Braunstein--Caves inequality increases as $s$ increases
\cite{BC88}. On the other hand, the growth of $s$ leads to
decreasing of a range of values $\theta$, for which violations
occur. The situation under consideration was also examined within
the $q$-entropic approach \cite{rastqic14}. With the above choice
of the vectors, the Bell type inequality of Ref. \cite{rastqic14}
is actually tantamount to (\ref{chshdq}). In this regard, the
second inequality (\ref{chshwq}) is a novel result. The
$q$-entropic approach allows to get some advances \cite{rastqic14}.
First, we can significantly expand a class of probability
distributions, for which the non-locality is testable in this way.
That is, some variations of $q\geq1$ allow to wide a range of
values $\theta$, for which violations occur. Second, the
$q$-entropic inequalities are expedient in analyzing cases with
detection inefficiencies \cite{rastqic14}. Two models of detection
inefficiencies in combination with the Braunstein--Caves
inequality were considered in Ref. \cite{rchtf12}.

Real experiments to test Bell inequalities are all non-ideal
\cite{shafi04}. Violations of the Bell inequalities in the
presence of decoherence and noise were studied in several papers
\cite{lixu04,lixu05,sclgx08,lsz1005,kofman12}. However,
information-theoretic formulations of the Bell theorem were not
addressed therein. Experimental studies of contextual properties
of polarization states of biphotons under decoherence are reported
in Ref. \cite{she14}. We shall now show that the $q$-entropic
approach is very useful in analyzing data of experiments in
decohering environment. Here, we will mainly focus on
(\ref{chshwq}), since it was not considered previously. Suppose
that each of two particles is subjected to dephasing noise during
its propagation from the point of emission to the detector. The
Kraus operators of the phase damping channel are written as
\cite{nielsen}
\begin{equation}
\ems_{0}=
\begin{pmatrix}
1 & 0 \\
0 & \sqrt{1-\lambda}
\end{pmatrix}
\, , \qquad
\ems_{1}=
\begin{pmatrix}
0 & 0 \\
0 & \sqrt{\lambda}
\end{pmatrix}
\, . \label{krop}
\end{equation}
For brevity, we denote $\lambda(t)=1-\exp(-2\gamma{t})$. During
the time $t$, the density matrix of a single qubit is mapped as
\begin{equation}
\bro\mapsto\cle(\bro)=
\ems_{0}{\,}\bro{\,}\ems_{0}+\ems_{1}{\,}\bro{\,}\ems_{1}
\ . \label{limap}
\end{equation}
This channel describes one of fundamental quantum effects. In
particular, it is helpful in understanding why a ``live-dead''
superposition of the Schr\"{o}dinger cat becomes unlikely. It is
also known that the phase damping channel can easily be converted
to the phase flip one by a simple recombination \cite{nielsen}.

If each of the two qubits is changed as (\ref{limap}), then the
initial state (\ref{ztsp}) is transformed to
\begin{equation}
(\cle\otimes\cle)
\bigl(|\Phi\rangle\langle\Phi|\bigr)=
\sum_{j,k=0}^{1}
(\ems_{j}\otimes\ems_{k})|\Phi\rangle\langle\Phi|(\ems_{j}\otimes\ems_{k})
\ . \label{mapli}
\end{equation}
By $\delta{t}_{1}$, we further mean the interval between the
emission and the first local measurement on $\cla$. The second
local measurement on $\clb$ will be performed $\delta{t}_{2}$
later. By calculations, we then obtain
\begin{equation}
(\cle_{1}\otimes\cle_{1})
\bigl(|\Phi\rangle\langle\Phi|\bigr)=
\exp(-2\gamma{\,}\delta{t}_{1})|\Phi\rangle\langle\Phi|
+\frac{\lambda(\delta{t}_{1})}{2}{\,}
\Bigl(
|0\rangle\langle0|\otimes|1\rangle\langle1|+
|1\rangle\langle1|\otimes|0\rangle\langle0|
\Bigr)
\> . \label{mapli1}
\end{equation}
Here, the subscript ``$1$'' marks that this map is related to the
first interval. Let us examine the case, when the plane of the
vectors $\vec{a}$, $\vec{b}^{\prime}$, $\vec{a}^{\prime}$,
$\vec{b}$ is orthogonal to the axis $z$. Dealing with
probabilities, we can rescale the considered observables. Suppose
that we first measure $\vec{a}\cdot\vec{\bsg}$ on the qubit
$\cla$. With the pre-measurement state (\ref{mapli1}), the
outcomes are equiprobable irrespectively to
$\gamma{\,}\delta{t}_{1}$. When the outcome $m$ has been obtained,
the post-measurement state is
\begin{equation}
p(m)^{-1}(\pisf_{m}(\vec{a})\otimes\pen_{2})
\left\{(\cle_{1}\otimes\cle_{1})
\bigl(|\Phi\rangle\langle\Phi|\bigr)\right\}
(\pisf_{m}(\vec{a})\otimes\pen_{2})
=\exp(-2\gamma{\,}\delta{t}_{1}){\>}\pisf_{m}(\vec{a})\otimes\pisf_{-m}(\vec{a})
+\lambda(\delta{t}_{1}){\>}\pisf_{m}(\vec{a})\otimes\bro_{*}
\ . \label{fpmst}
\end{equation}
By $\pisf_{m}(\vec{a})$, we denote the orthogonal projector on the
eigenstate of $\vec{a}\cdot\vec{\bsg}$ corresponding to the
eigenvalue $m$. By $\bro_{*}=\pen_{2}/2$, we mean the completely
mixed state of a qubit. The post-first-measurement state is
further mapped by the phase damping channel during the interval
$\delta{t}_{2}$. Since the state (\ref{fpmst}) is separable, it is
merely mapped to
\begin{equation}
\exp(-2\gamma{\,}\delta{t}_{1}){\>}\cle_{2}\bigl(\pisf_{m}(\vec{a})\bigr)\otimes\cle_{2}\bigl(\pisf_{-m}(\vec{a})\bigr)
+\lambda(\delta{t}_{1}){\>}\cle_{2}\bigl(\pisf_{m}(\vec{a})\bigr)\otimes\bro_{*}
\ . \label{efpmst}
\end{equation}
Here, the subscript ``$2$'' marks that this map is related to the
second interval. We also used that the completely mixed state is a
fixed point the phase damping channel. The pre-measurement state
of the qubit $\clb$ is obtained by the partial trace operation:
\begin{equation}
\exp(-2\gamma{\,}\delta{t}_{1}){\>}\cle_{2}\bigl(\pisf_{-m}(\vec{a})\bigr)
+\lambda(\delta{t}_{1}){\,}\bro_{*}
=\frac{1}{2}{\,}
\bigl(
\pen_{2}+\vec{v}\cdot\vec{\bsg}
\bigr)
\ . \label{fepmst}
\end{equation}
That is, the pre-measurement density matrix is represented by its
Bloch vector $\vec{v}=(v_{x},v_{y},v_{z})$. In terms of the Bloch
vector, one gets
\begin{equation}
\vec{v}=-m\exp(-2\gamma{\,}\delta{t}_{1})\exp(-\gamma{\,}\delta{t}_{2}){\>}\vec{a}
\> . \label{blcv}
\end{equation}
The action of the map $\cle_{2}$ shrinks horizontal components of
the initial Bloch vector by the factor
$\exp(-\gamma{\,}\delta{t}_{2})=\sqrt{1-\lambda(\delta{t}_{2})}$.
Measuring the observable $\vec{b}\cdot\vec{\bsg}$, the outcome
$m^{\prime}$ occurs with the probability
\begin{equation}
\frac{1}{2}{\,}\bigl(1+m^{\prime}\vec{b}\cdot\vec{v}\bigr)
=\frac{1-m^{\prime}m\exp(-\gamma{\,}\delta{t})\cos\theta}{2}=p(B=m^{\prime}|A=m)
\ . \label{bmpam}
\end{equation}
Here, we use (\ref{angls}) and denote
$\delta{t}=2{\,}\delta{t}_{1}+\delta{t}_{2}$. Thus, we obtain the
conditional probability $p(B=m^{\prime}|A=m)$. It turns out that
the final expression (\ref{bmpam}) is symmetric in the labels $m$
and $m^{\prime}$. For other pairs of jointly measurable
observables of interest, conditional probabilities are obtained by
replacing $\theta$ with $\theta/3$ in the formula (\ref{bmpam}).

\begin{figure}
\includegraphics[width=9.0cm]{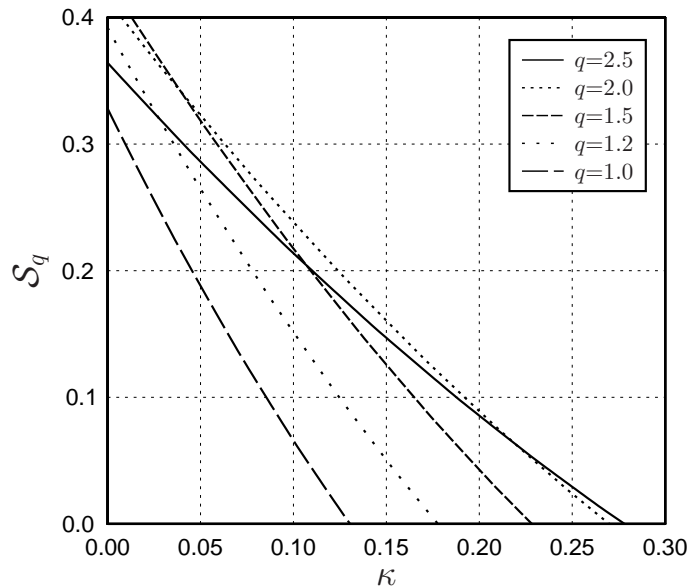}
\caption{\label{fig1}The dependence $\mathcal{S}_{q}(\kappa)$ in
the CHSH scenario with noise for five values of $q$, namely
$q=1.0;1.2;1.5;2.0;2.5$. For each value of $q$, only positive
values of $\mathcal{S}_{q}$ are shown.}
\end{figure}

We shall now show that the obtained probability distributions
sometimes violate the locality conditions (\ref{chshdq}) and
(\ref{chshwq}). To characterize a violation of the restriction
(\ref{chshwq}), we introduce a characteristic quantity
\begin{align}
\mathcal{C}_{q}:=\dw_{q}(A,B)-\dw_{q}(A,B^{\prime})-\dw_{q}(B^{\prime},A^{\prime})-\dw_{q}(A^{\prime},B)
\ . \label{chshdw}
\end{align}
Strictly positive values of (\ref{chshdw}) will reveal violations
of the $q$-metric inequality (\ref{chshwq}) in the presence of
decoherence. A possibility to detect such violations essentially
depends on the entropic parameter $q$. Note also that a strength
of violations depends on values of the parameter
$\gamma{\,}\delta{t}$. This parameter characterizes the influence
of an environment in the model considered. To study the question,
we put the ratio
\begin{equation}
\kappa:=\frac{\gamma{\,}\delta{t}}{\theta/3}
\ . \label{vkdf0}
\end{equation}
This quantity linearly increases with $\gamma$ as well as with
$\delta{t}$. The characteristic quantity (\ref{chshdw}) is some
function $\mathcal{C}_{q}(\theta,\kappa)$ of two variables. We
would like to see a trade-off between $\kappa$ and $q$. Let us
define the quantity
\begin{equation}
\mathcal{S}_{q}(\kappa):=\underset{\theta}{\sup}{\>}\mathcal{C}_{q}(\theta,\kappa)
\ . \label{ovmc}
\end{equation}
Such an approach is meaningful, since in real experiments we want
to maximize a possible violation to be tested. It is very useful
that the range of positivity of (\ref{ovmc}) essentially depends
on $q$. To be more precise, we introduce the bound
\begin{equation}
\kappa_{s}(q)=\sup\!\left\{\kappa:{\,}\kappa\geq0,{\>}\mathcal{S}_{q}(\kappa)>0\right\}
\, . \label{bkap}
\end{equation}
The first fact is that the strength of violations essentially
depends on the parameter $q\geq1$. To consider a behavior of the
range of violations, we focus on $\kappa_{s}(q)$. It turns out
that $\kappa_{s}(q)$ increases with $q$. The dependence of
$\mathcal{S}_{q}(\kappa)$ on $\kappa$ is
shown in Fig. \ref{fig1}. For the first time, both the strength
and the range of positivity are increased with $q$. Indeed, the
curves with $q>1$ all lie over the curve for $q=1$. For
sufficiently large $q$, however, the strength of positivity
becomes reducing. Nevertheless, the least point $\kappa_{s}(q)$
still slowly increases with growing $q$. These results
maintain a conclusion that the inequalities in terms of Tsallis'
entropies give a suitable tool in studying the noisy case.

The above scheme could be used in cases, when observations have
more than two outcomes. Of course, calculations become more
complicated as well. From the experimental viewpoint, the
trichotomic case is also important. Indeed, the CHSH setup with
trichotomic observables can be realized with a pair of biphotons
\cite{kwek02}. A violation of the $q$-metric inequalities in the
CHSH case with trichotomic outcomes is also more robust to
decoherence by varying the parameter. We refrain from presenting
the details here, since similar conclusions were found in Ref.
\cite{mkk14}. Instead, we will address the noisy trichotomic case
in the Leggett--Garg scenario. The authors of Ref. \cite{mkk14}
modeled a noise by adding a completely mixed term to the
noise-free density matrix of qutrit pair. We used more detailed
approach, in which the environmental influence is taken into
account through the phase damping channel applied to each of the
particles.

Thus, the $q$-entropic approach is essential in analyzing data of
Bell-type experiments in dephasing environments. In some cases,
formulation with generalized entropies may be compared with the
following approach. Adding a shared randomness into the
experiment, the Shannon-entropy inequalities sometimes give the
full information to conclude if a given correlation is non-local
or not \cite{rch13}. By the depolarization protocol of Ref.
\cite{mag06}, this result was shown for any $n$-cycle with
dichotomic outcomes \cite{rch13}. However, experimental setups
with a shared randomness may be vulnerable to noise. A way to
implement such setting in the Leggett--Garg scenario is not
obvious. These questions could be a theme of separate research.
Further, the $n$-cycle scenario with trichotomic and more outcomes
is also of interest. The use of $q$-entropies extended not only a
class of probability distributions, for which incompatibility with
the local realism or the macrorealism is testable in entropic
terms \cite{rastqic14,rastq14}. Additionally, the $q$-entropic
approach is further motivated by its advantages in studying
experiments with detection inefficiencies and by its robustness in
the noisy case.

\section{On metric Leggett--Garg inequalities under decoherence}\label{sec4}

In this section, we will deal with information-theoretic
inequalities of the Leggett--Garg type. These inequalities are
based on the following two concepts \cite{eln13}. First, we assume
that physical properties of a macroscopic object preexist
irrespectively to the act of observation. Second, measurements are
non-invasive in the sense that the measurement of an observable at
any instant of time does not alert its subsequent evolution.
Following the ideas of Ref. \cite{uksr12}, we consider the
two-level system with the self Hamiltonian
\begin{equation}
\hm=-\omega\sm_{z}=-\frac{\hbar\omega}{2}{\>}\bsg_{z}
\ . \label{sham}
\end{equation}
Its eigenstates are the ground state $|0\rangle$ with energy
$-\hbar\omega/2$ and the excited state $|1\rangle$ with energy
$+\hbar\omega/2$. These states are explicitly written as
(\ref{01bs}). In studies of restrictions of the Leggett--Garg
type, the Heisenberg picture is more convenient. The operator of
unitary evolution is
\begin{equation}
\um(t)=\exp(-\iu\hbar^{-1}t\hm)
=\exp\bigl(+\iu(\omega{t}/2)\bsg_{z}\bigr)
\ . \label{vam}
\end{equation}
For brevity, we refer to measured results of some spin component as
$\pm1$. Let us study information-theoretic inequalities of the
Leggett--Garg type for the $x$-component of the spin. Assuming a
validity of the macrorealistic approach, we will deal with the
quantity $S_{x}(t)$. Our description of $S_{x}(t)$ should be
carried out in line with the macroscopic realism {\it per se} and
the non-invasive measurability. Let $\tau$, $\tau^{\prime}$,
$\tau^{\prime\prime}$ be three instants of the time. Denoting
$X=S_{x}(\tau)$, $X^{\prime}=S_{x}(\tau^{\prime})$,
$X^{\prime\prime}=S_{x}(\tau^{\prime\prime})$, for $q\geq1$ we
write the conditions
\begin{align}
\Delta_{q}(X,X^{\prime\prime})&\leq\Delta_{q}(X,X^{\prime})+\Delta_{q}(X^{\prime},X^{\prime\prime})
\ , \label{sthw}\\
\dw_{q}(X,X^{\prime\prime})&\leq\dw_{q}(X,X^{\prime})+\dw_{q}(X^{\prime},X^{\prime\prime})
\ . \label{stdw}
\end{align}
These formulas give a formulation of the Leggett--Garg
inequalities in terms of the metrics (\ref{dqxy}) and
(\ref{dwqxy}), respectively. The inequalities (\ref{sthw}) and
(\ref{stdw}) are sometimes violated by probability distributions
calculated in quantum-mechanical way. The initial state is chosen
to be completely mixed \cite{uksr12}. In the basis
$\bigl\{|0\rangle,|1\rangle\bigr\}$, eigenstates of the operator
$\sm_{x}=(\hbar/2)\bsg_{x}$ are written as
\begin{equation}
|x_{\pm1}\rangle=
\frac{1}{\sqrt{2}}
\begin{pmatrix}
1 \\
\pm1
\end{pmatrix}
\, . \label{xpmdf}
\end{equation}
In the Heisenberg picture, we deal with the operator
$\um(t)^{\dagger}{\,}\sm_{x}{\,}\um(t)$. It describes the
evolution of the $x$-component of the spin. The aim is to obtain
the corresponding conditional probabilities. They will show a
violation of the inequalities (\ref{sthw}) and (\ref{stdw}) with a
concrete example of spin-1/2 particle.

If the system is not altered by the environment, then its initial
state $\bro_{0}$ remains unchanged up to measurement.
When we measure the $x$-component at the moment $t=\tau$, the
outcome $m=\pm1$ occurs with probability
\begin{equation}
p(m)=\Tr\bigl(\mpi_{m}(\tau)\bro_{0}\bigr)
\> , \label{pmex}
\end{equation}
where the corresponding projector
\begin{equation}
\mpi_{m}(\tau)=\um(\tau)^{\dagger}|x_{m}\rangle\langle{x}_{m}|\um(\tau)
\ . \label{pvdf}
\end{equation}
The density matrix
$p(m)^{-1}\mpi_{m}(\tau){\,}\bro_{0}{\,}\mpi_{m}(\tau)$ describes
the corresponding post-measurement state. For the initial state
$\bro_{0}=\pen_{2}/2$, we get $p(m)=1/2$ and the post-measurement
state $\mpi_{m}(\tau)$. Then the conditional probability of
obtaining the outcome $m^{\prime}$ at the next time
$t=\tau^{\prime}$ is equal to
\begin{equation}
p(m^{\prime}|m)=
\Tr\bigl(\mpi_{m^{\prime}}(\tau^{\prime})\mpi_{m}(\tau)\bigr)=
\left|
\langle{x}_{m^{\prime}}|\um(\tau^{\prime})
\um(\tau)^{\dagger}|x_{m}\rangle
\right|^{2}
\> . \label{mpmwj}
\end{equation}
The right-hand side of (\ref{mpmwj}) is immediately connected with
elements of the corresponding rotation matrix. Such matrices are
well studied \cite{bl81}. In the case considered, we obtain the
probabilities
\begin{equation}
p(m^{\prime}|m)=
\frac{1+m^{\prime}m\cos\omega(\tau^{\prime}-\tau)}{2}
\ , \label{mpmwj1}
\end{equation}
where $m,m^{\prime}=\pm1$. As was shown in Ref.
\cite{uksr12}, conditional probabilities of the form
(\ref{mpmwj1}) can violate entropic inequalities of the
Leggett--Garg type. In real experiments, however, quantum systems
are inevitably exposed to noise. We shall theoretically study
possibilities to test a violation of the macrorealistic picture in
the presence of decoherence.

Suppose that the qubit is exposed to noise. For all operators of
interest, we will therefore consider a transformation of the form
\begin{equation}
\xm\mapsto\um(t)^{\dagger}{\,}\xm{\,}\um(t)
\ . \label{trun}
\end{equation}
In the noisy case, the transformation (\ref{trun}) implies the use
of the interaction picture instead of the Heisenberg one. The
density matrix denoted by $\bro_{I}$ now corresponds to the
interaction picture and changes during the evolution. These
changes are fully related to the environmental influence. Like the
analysis of the previous section, we can assume that the density
matrix is mapped by some quantum channel. In terms of the Bloch
vector, we represent $\bro_{I}$ as
\begin{equation}
\bro_{I}=\frac{1}{2}{\,}
\bigl(
\pen_{2}+\vec{w}\cdot\vec{\bsg}
\bigr)
\, . \label{blch}
\end{equation}
Any qubit channel can be represented by its action on $\vec{w}$.

Another way to describe the above situation is the use of quantum
master equation \cite{petr2002}. It is generally written in the
Kossakowski--Lindblad form \cite{akoss72,lind76}
\begin{equation}
\frac{\xdif}{\xdif{t}}{\>}\bro_{S}(t)=
-\frac{\iu}{\hbar}{\>}[\hm,\bro_{S}]+\sum_{j}\gamma_{j}
\left(\am_{j}\bro_{S}\am_{j}^{\dagger}-\frac{1}{2}{\>}\am_{j}^{\dagger}\am_{j}\bro_{S}
-\frac{1}{2}{\>}\bro_{S}\am_{j}^{\dagger}\am_{j}
\right)
\, . \label{lbeq}
\end{equation}
The index ``$S$'' marks that the density matrix is related to the
Schr\"{o}dinger picture. The operators $\am_{j}$ are usually
referred to as the Lindblad operators \cite{petr2002}. The
parameters $\gamma_{j}$ play the role of relaxation rates for
different decay modes of the open system. Concrete kinds of
environment influence are reflected by the corresponding Lindblad
operators. When the system evolution is described by (\ref{lbeq}),
the transformation (\ref{trun}) will imply replacing the
Schr\"{o}dinger picture by the interaction picture. We will
consider two important models of decohering processes described by
the phase damping and depolarizing channels.

Let $\delta\tau$ be the time interval between two successive
measurements of the $x$-component of the spin. We first suppose
that the Bloch vector is mapped as
\begin{equation}
(w_{x},w_{y},w_{z})\longmapsto
\bigl(\sqrt{1-\lambda}{\,}w_{x},\sqrt{1-\lambda}{\,}w_{y},w_{z}\bigr)
\ , \label{wlaw}
\end{equation}
where $\lambda(\delta\tau)=1-\exp(-2\gamma{\,}\delta\tau)$. This
discrete transformation of the Bloch vector corresponds to the
phase damping channel \cite{nielsen}. To put the transformation
(\ref{wlaw}) into the form (\ref{lbeq}), we introduce the operator
\begin{equation}
\nam:=|1\rangle\langle1|
\ . \label{lamdf}
\end{equation}
It shows the number of excitations, since
$\nam{\,}|0\rangle=0{\,}|0\rangle$ and
$\nam{\,}|1\rangle=1{\,}|1\rangle$. The qubit in dephasing
environment can be described by the quantum master equation
\begin{equation}
\frac{\xdif}{\xdif{t}}{\>}\bro_{S}(t)=
-\frac{\iu}{\hbar}{\>}[\hm,\bro_{S}]+\gamma
\bigl(2\nam\bro_{S}\nam-\nam\bro_{S}
-\bro_{S}\nam
\bigr)
\ . \label{lbphd}
\end{equation}
Here, we take into account that the operator (\ref{lamdf}) is
Hermitian and projective, i.e., $\nam^{2}=\nam$. We also note that
the operator (\ref{lamdf}) commutes with $\bsg_{z}$ and is
invariant under the transformation (\ref{trun}). For a
convenience in consequent expressions, the parameter $\gamma$ is
rescaled as well. In the interaction picture, the corresponding
density matrix
\begin{equation}
\bro_{I}(t)=\um(t)^{\dagger}\bro_{S}(t){\,}\um(t)
\label{dminp}
\end{equation}
is changed according to the equation
\begin{equation}
\frac{\xdif}{\xdif{t}}{\>}\bro_{I}(t)=
\gamma
\bigl(
2\nam\bro_{I}\nam-\nam\bro_{I}-\bro_{I}\nam
\bigr)
\ . \label{lbphdi}
\end{equation}
Substituting (\ref{blch}) to (\ref{lbphdi}) finally gives
$\overset{\centerdot}{w}_{x}=-\gamma{w}_{x}$,
$\overset{\centerdot}{w}_{y}=-\gamma{w}_{y}$, and
$\overset{\centerdot}{w}_{z}=0$. So, after the time $t$ the
initial horizontal components are multiplied by
$\exp(-\gamma{t})$; the $z$-component remains constant. In other
words, off-diagonal elements of the density matrix (\ref{blch})
are exponentially decayed. By the phase damping channel, the Bloch
ball is turned into an ellipsoid touching the Bloch sphere at the
north and south poles \cite{bengtsson}.

We shall now recalculate the conditional probabilities
(\ref{mpmwj1}) in the presence of dephasing environment. During
the time interval from the first to the second measurements,
components of the Bloch vector of any post-first-measurement state
are changed as follows. First, a unitary transformation with the
generator $\hm$ rotates the Bloch vector around the $z$-axis.
Second, the phase damping rescales horizontal components by the
factor $\exp(-\gamma{\,}\delta\tau)$, where
$\delta\tau=\tau^{\prime}-\tau$. More precisely, in the
interaction picture we write
\begin{equation}
p(m^{\prime}|m)=
\Tr\bigl(\mpi_{m^{\prime}}(\tau^{\prime})\bro_{Im}(\tau^{\prime})\bigr)
\> . \label{mpmwp}
\end{equation}
Here, the density matrix $\bro_{Im}(\tau^{\prime})$ is obtained
from the post-first-measurement state $\mpi_{m}(\tau)$ according
to the phase damping with the factor
$\exp(-\gamma{\,}\delta\tau)$. It follows from (\ref{pvdf}) that
\begin{equation}
\bro_{Im}(\tau^{\prime})=
\frac{1}{2}{\,}
\Bigl(
\pen_{2}+\exp(-\gamma{\,}\delta\tau){\,}m{\,}\um(\tau)^{\dagger}\bsg_{x}\um(\tau)
\Bigr)
\> . \label{mpmpw}
\end{equation}
Similarly to (\ref{mpmwj1}), we then get the final expression
\begin{equation}
p(m^{\prime}|m)=
\frac{1+m^{\prime}m\exp(-\gamma{\,}\delta\tau)\cos\omega\delta\tau}{2}
\ . \label{mpmwj2}
\end{equation}
Thus, the influence of dephasing environment merely results in
exponential decay of non-trivial terms of conditional
probabilities.

The completely mixed state is a fixed point for a unitary
evolution as well as for the phase damping channel. Indeed, the
phase damping channel is unital. The role of unitality against
unitarity in the context of quantum fluctuation theorems was
recently revealed \cite{albash,rast13,kz14}. Unital channels with
controlled amount of noise were used in experimental studies of
the non-local and contextual properties of biphotons \cite{she14}.
Following Refs. \cite{uksr12,rastq14}, we consider three
measurements in equidistant time intervals. Measuring the
$x$-component at the moment $t=\tau$, we will again have outcomes
$m=\pm1$ with the probability $p(m)=1/2$. The latter is
conditioned by the choice of the initial state. Taking $p(m)$ and
$p(m^{\prime}|m)$, we obtain the joint probabilities
$p(m,m^{\prime})$ of the outcomes $m$ at $t=\tau$ and $m^{\prime}$
at $t=\tau^{\prime}$. We also note that the expression
(\ref{mpmwj2}) is symmetric with respect to the labels $m$ and
$m^{\prime}$. These points allow to evaluate the corresponding
distance. Instead of (\ref{chshdw}), the characteristic quantity
is now expressed as
\begin{align}
\mathcal{C}_{q}:=\dw_{q}(X,X^{\prime\prime})-\dw_{q}(X,X^{\prime})-\dw_{q}(X^{\prime},X^{\prime\prime})
\ . \label{mcdw}
\end{align}
In (\ref{mcdw}), the distance $\dw_{q}(X,X^{\prime})$ is found
from the probabilities $p(m,m^{\prime})$. The distances
$\dw_{q}(X^{\prime},X^{\prime\prime})$ and
$\dw_{q}(X,X^{\prime\prime})$ are obtained in the same manner.
Here, the conditional probabilities $p(m|m^{\prime\prime})$ are
expressed like (\ref{mpmwj2}), but with the interval $2{\,}\delta\tau$
instead of $\delta\tau$.

It is natural that an influence of the phase damping process is
dependent on the ratio of its rate to the excitation frequency. To
study this question, we introduce an analog of (\ref{vkdf0})
written as $\kappa:=\gamma/\omega$. In the notation of
(\ref{lbphd}), the parameter $\gamma$ is taken to be a half of the
relaxation rate. Of course, the quantity (\ref{mcdw}) also depends
on the entropic parameter $q$. It is convenient to put an
auxiliary variable $\theta=\omega\delta\tau$. This substitution
will allow us to exploit a similarity between the conditional
probabilities (\ref{bmpam}) and (\ref{mpmwj2}). The characteristic
quantity (\ref{mcdw}) then becomes some function
$\mathcal{C}_{q}(\theta,\kappa)$ of two variables. In the previous
section, the angle $\theta$ was a characteristic of the geometry
of experiment. In this section, however, the treatment of $\theta$
is purely temporal. For the given frequency $\omega$, values of
the variable $\theta$ can be controlled by choosing $\delta\tau$.

Focusing on a behavior with respect to $\kappa$, we will again
take the optimization of $\mathcal{C}_{q}(\theta,\kappa)$ over
$\theta$ for the fixed $\kappa$ and $q$. Formally, the function
$\mathcal{S}_{q}(\kappa)$ is again defined by (\ref{ovmc}). The
only change is that the term $\mathcal{C}_{q}$ is defined by
(\ref{mcdw}) instead of (\ref{chshdw}). Strictly positive values
of the quantity (\ref{ovmc}) will reveal a violation of the
Leggett--Garg restrictions in the presence of dephasing
environment. Additional ways of analysis of experimental data are
provided by a possibility to vary the entropic parameter $q$. As
was shown, the $q$-entropic approach can allow to reduce an amount
of required detection efficiency \cite{rastqic14,rastq14}. We
shall now motivate that a possibility to vary $q$ is also
significant from the viewpoint of analyzing data of experiments in
the presence of decoherence.

\begin{figure}
\includegraphics[width=9.0cm]{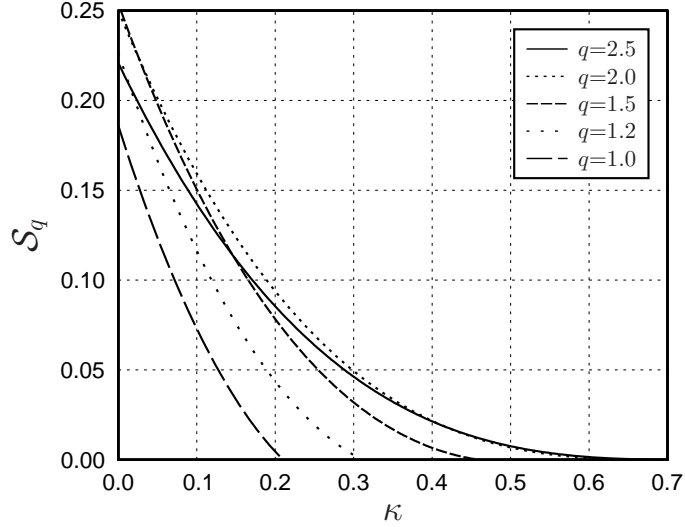}
\caption{\label{fig2}The dependence $\mathcal{S}_{q}(\kappa)$ in
the Leggett--Garg scenario with spin-$1/2$ system for five values
of $q$, namely $q=1.0;1.2;1.5;2.0;2.5$. For each value of $q$,
only positive values of $\mathcal{S}_{q}$ are shown.}
\end{figure}

It is instructive to discuss the quantity $\kappa_{s}(q)$
introduced by (\ref{bkap}) as well. It is important that the range
$[0;\kappa_{s}(q)]$ essentially depends on $q\geq1$. In Fig.
\ref{fig2}, we have shown $\mathcal{S}_{q}(\kappa)$ versus
$\kappa$ for several values of $q$, including the standard case
$q=1$. With growing $q$, both the strength and the range of
positivity are firstly increased. In effect, the curve with $q>1$
all go over the curve for $q=1$. With growing $q>1$, the point
$\kappa_{s}(q)$ also increases. For sufficiently large $q$,
however, the strength of positivity becomes reducing.
Nevertheless, the least point $\kappa_{s}(q)$ is still slowly
increasing with growth of $q$. In principle, we can actually
restrict a consideration to values of $q$ around the point
$q=2.0$.

We now examine another relaxation process, which is described with
the Lindblad operator $\am_{j}=\bsg_{j}$ for $j=x,y,z$. One leads
to the master equation
\begin{equation}
\frac{\xdif}{\xdif{t}}{\>}\bro_{S}(t)=
-\frac{\iu}{\hbar}{\>}[\hm,\bro_{S}]+\gamma\sum_{j=x,y,z}
\left(\bsg_{j}{\,}\bro_{S}{\,}\bsg_{j}-\bro_{S}
\right)
\, . \label{eqlb}
\end{equation}
Using the unitary transformation (\ref{trun}), we then rewrite
(\ref{eqlb}) in the interaction picture. The way is quite similar
to the pass from (\ref{lbphd}) to (\ref{lbphdi}). We refrain from
presenting the details here. The following result takes place. Due
to decoherence, the three components the Bloch vector of
$\bro_{I}$ are all multiplied by the factor $\exp(-4\gamma{t})$.
In other words, the Bloch vector is transformed as
\begin{equation}
(w_{x},w_{y},w_{z})\longmapsto
\bigl((1-4\mu/3)w_{x},(1-4\mu/3)w_{y},(1-4\mu/3)w_{z}\bigr)
\> , \label{laww}
\end{equation}
where $\mu(t)=(3/4)\bigl(1-\exp(-4\gamma{t})\bigr)$. The
transformation (\ref{laww}) corresponds to the depolarizing
channel with the four Kraus operators $\sqrt{1-\mu}{\,}\pen$ and
$\sqrt{\mu/3}{\,}\bsg_{j}$ for $j=x,y,z$. This channel merely
shrinks the Bloch ball \cite{bengtsson}.

Instead of shrinking of horizontal components of the Bloch vector,
we now deal with shrinking of the vector as a whole. As was
discussed right before (\ref{mpmwj2}), conditional probabilities
of interest are determined by changes of horizontal components of
the Bloch vector. Recalculating the conditional probabilities
(\ref{mpmwj1}), we merely replace the expression (\ref{mpmwj2}) by
\begin{equation}
p(m^{\prime}|m)=
\frac{1+m^{\prime}m\exp(-4\gamma{\,}\delta\tau)\cos\omega\delta\tau}{2}
\ . \label{mpmwj22}
\end{equation}
The only distinction is that decaying of non-trivial terms in
conditional probabilities is much faster than in (\ref{mpmwj2}).
Thus, the above conclusions can all be applied to the case of
depolarizing environment. We should only rescale the ratio $\kappa$
appropriately.

Our conclusions remain valid for trichotomic observables in the
Leggett--Garg scenario with noise. The spin-$1$ observables are
represented as matrices
\begin{equation}
\sm_{x}=\frac{\hbar}{\sqrt{2}}
\begin{pmatrix}
0 & 1 & 0 \\
1 & 0 & 1 \\
0 & 1 & 0
\end{pmatrix}
, \qquad
\sm_{y}=\frac{\hbar}{\sqrt{2}}
\begin{pmatrix}
0 & -\iu & 0 \\
\iu & 0 & -\iu \\
0 & \iu & 0
\end{pmatrix}
, \label{jxy}
\end{equation}
and $\sm_{z}=\hbar{\,}\textup{diag}(+1,0,-1)$. Each of these
matrices has eigenvalues $0$, $\pm\hbar$. Further, we rescale the
eigenvalues as $m\in\{+1,0,-1\}$. The noise-free evolution of the
$x$-component is governed as
$\um(t)^{\dagger}{\,}\sm_{x}{\,}\um(t)$ with substituting the
corresponding spin-$1$ matrices. Instead of (\ref{sham}), the self
Hamiltonian is now represented as
\begin{equation}
\hm=-\hbar\omega{\>}{\textup{diag}}(+1,0,-1)
\ . \label{sham1}
\end{equation}
The energy levels are $-\hbar\omega$, $0$, $+\hbar\omega$.
Dephasing processes will be described as follows. Let density
matrix in the interaction picture be changed according to the
master equation
\begin{equation}
\frac{\xdif}{\xdif{t}}{\>}\bro_{I}(t)=\bsl[\bro_{I}]
\ . \label{meq}
\end{equation}
Generating evolution in the interaction picture, the linear
superoperator is written as
\begin{equation}
\bsl[\bro_{I}]=\gamma
\bigl(
2\nam\bro_{I}\nam-\nam^{2}\bro_{I}-\bro_{I}\nam^{2}
\bigr)
\, , \label{gds}
\end{equation}
where $\nam={\textup{diag}}(-1,0,+1)$. The latter is not changed
by (\ref{trun}). Like the case of spin-$1/2$, the operator $\nam$
is related to a number of excitations. It is calculated with
respect to the zero energy level. However, the matrix
${\textup{diag}}(-1,0,+1)$ is not idempotent. Except for this
fact, the master equation (\ref{meq}) is fully similar to
(\ref{lbphdi}).

It is convenient to represent a density matrix in terms of
generalized Bloch vector \cite{bengtsson}. Let $\blm_{k}$'s denote
the standard Gell-Mann matrices. Instead of (\ref{blch}), we now
write
\begin{equation}
\bro_{I}=\frac{1}{3}
\left(
\pen_{3}+\sum\nolimits_{k=1}^{8}w_{k}{\,}\blm_{k}
\right)
\, , \qquad
w_{k}=\frac{3}{2}{\,}\Tr(\bro_{I}\blm_{k})
\ . \label{rhorp}
\end{equation}
Calculations give $\bsl[\blm_{3}]=\bsl[\blm_{8}]=\bnil$,
$\bsl[\blm_{k}]=-\gamma\blm_{k}$ for $k=1,2,6,7$, and
$\bsl[\blm_{k}]=-4\gamma\blm_{k}$ for $k=4,5$. That is, the vector
components $w_{3}$ and $w_{8}$ are constant, whereas other ones
are exponentially decayed. So, the off-diagonal terms of a density
matrix decay like either $\exp(-\gamma{t})$ or
$\exp(-4\gamma{t})$. Such a phase decoherence takes place between
the moments of observations.

\begin{figure}
\includegraphics[width=9.0cm]{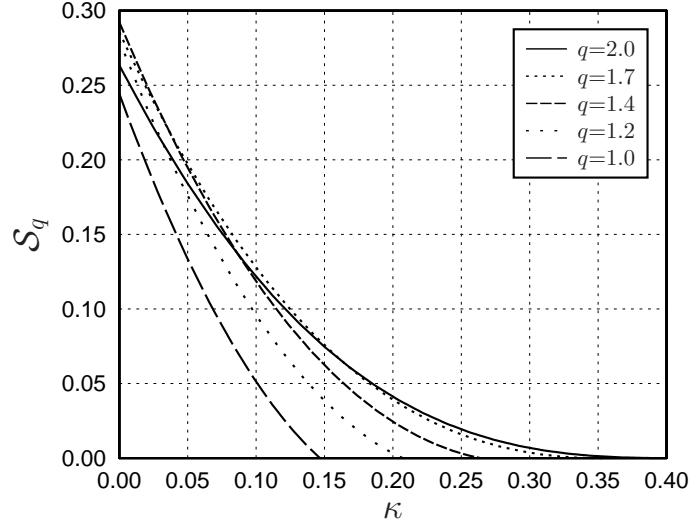}
\caption{\label{fig3}The dependence $\mathcal{S}_{q}(\kappa)$ in
the Leggett--Garg scenario with spin-$1$ system for five values
of $q$, namely $q=1.0;1.2;1.4;1.7;2.0$. For each value of $q$,
only positive values of $\mathcal{S}_{q}$ are shown.}
\end{figure}

If the system is initially prepared in the state
$\bro_{0}=\pen_{3}/3$, then the distribution $p(m)$ is uniform.
The completely mixed state is a fixed point of phase damping.
Conditional probabilities are calculated similarly to the above
case. For all $m,m^{\prime}\in\{-1,0,+1\}$, they satisfy
$p(m^{\prime}|m)=p(m|m^{\prime})$. This property reflects the fact
that phase damping is unital. For outcomes $m,m^{\prime}=\pm1$,
one gets
\begin{equation}
p(m^{\prime}|m)=
\frac{3}{8}+\frac{m^{\prime}m}{2}{\>}\exp(-\gamma{\,}\delta\tau)\cos\omega\delta\tau
+\frac{1}{8}{\>}\exp(-4\gamma{\,}\delta\tau)\cos2\omega\delta\tau
\ . \label{mplg1}
\end{equation}
When one of the outcomes is $0$, we have
\begin{align}
p(\pm1|0)&=\frac{1-\exp(-4\gamma{\,}\delta\tau)\cos2\omega\delta\tau}{4}
\ , \label{m0lg1}\\
p(0|0)&=\frac{1+\exp(-4\gamma{\,}\delta\tau)\cos2\omega\delta\tau}{2}
\ . \label{00lg1}
\end{align}
For $\gamma=0$, the above probabilities lead to the squares of the
corresponding small Wigner $d$-functions written with
$\theta=\omega\delta\tau$. For the Leggett--Garg scenario without
noise, conditional probabilities in terms of $d$-functions were
given in Ref. \cite{uksr12}. In the case of equidistant time
intervals, the conditional probabilities $p(m|m^{\prime\prime})$
are expressed like (\ref{mplg1})--(\ref{00lg1}). We should only
replace $\delta\tau$ with $2{\,}\delta\tau$.

Keeping the corresponding probabilities, we further obtain
$q$-entropic distances in the right-hand side of (\ref{mcdw}).
Calculations show the existence of positive values of
$\mathcal{C}_{q}$ for trichotomic observables in the Leggett--Garg
scenario with noise. We again use $\mathcal{S}_{q}(\kappa)$ for
describing an amount of violation of the metric Leggett--Garg
inequality (\ref{stdw}). Recall that the quantity
$\mathcal{S}_{q}(\kappa)$ is formally posed by (\ref{ovmc}). This
quantity is shown in Fig. \ref{fig3} for several values of $q$.
The conclusions found above for dichotomic observables are all
actual for trichotomic observables. With growing $q$, both the
strength and the range of positivity are firstly increased.
Comparing Fig. \ref{fig3} with Fig. \ref{fig2}, we only note that
the range of positivity on the $\kappa$-axis is less for the
spin-$1$ system. For $\kappa=0$, values of $\mathcal{S}_{q}$ are
slightly larger for $s=1$. In units of $\ln_{q}(2s+1)$, however, the
strength of violation decreases with growth of $s$ \cite{rastq14,uksr12}.
Thus, a violation of the $q$-entropic inequality (\ref{stdw}) can
be made more robust to decoherence by adopting the parameter
value. We hope that the presented results could be useful in
analysis of real experiments to test the Leggett--Garg
inequalities.

\section{Conclusion}\label{sec5}

We examined capabilities of conditional entropies of the Tsallis
type in studying restrictions of the Bell type in the presence of
decoherence. One of unifying approaches to problems of local
realism and non-contextuality is provided by the so-called
triangle principle. The mentioned questions are related to
statistical predictions and, therefore, deal with probability
distributions. Any use of the triangle principle is based on some
metric in the probability space of interest. There are several
realizations of the triangle principle with the use of conditional
$q$-entropies of the Tsallis type. It turned out that both the
known forms of the conditional $q$-entropy lead to a legitimate
metric. Such metrics are naturally treated as an
information-theoretic distance between random variables related to
any pair of measurements. It seems that the metric based on the
second conditional form was previously not considered in the
literature. Using the defined metrics, we obtained the two
families of information-theoretic Bell inequalities, which depends
on one entropic parameter. We further considered
information-theoretic inequalities of the Bell and Leggett--Garg
types under decoherence. This question is studied in detail for
dichotomic and trichotomic spin observables in the presence of
noise environment. We considered a situation, when each of two
particles in the CHSH scenario is independently subjected to the
phase damping channel. As calculations showed, a violation of the
corresponding $q$-entropic inequality become much more robust to
decoherence by adopting the parameter value. Similar conclusions
were made in the taken example of $q$-entropic restrictions of the
Leggett--Garg type. A dynamics of this system is modeled by a
master equation written in the Kossakowski--Lindblad form. In the
interaction picture, a relaxation of the corresponding density
matrix is described by the phase damping or depolarizing channels.
It is natural that too fast relaxation will actually prefer tests
of the Leggett--Garg inequalities in real experiments. At the same
time, both the strength and range of violations can be increased
by adopting suitable values of the entropic parameter. For the
given ratio of relaxation rate to the excitation frequency, the
violation may still be testable with certain values of $q$. Of
course, a rate of relaxation process should be low enough. Thus,
the $q$-entropic formulation could be useful in the analysis of recent
experiments to test the contextual and non-local properties.

\end{document}